# Graph Neural Networks for an Accurate and Interpretable Prediction of the Properties of Polycrystalline Materials

Minyi Dai,[1‡] Mehmet F. Demirel,[2‡] Yingyu Liang,[2†] Jia-Mian Hu[1*]

*[1]Department of Materials Science and Engineering, University of Wisconsin-Madison, WI, 53706*

*[2]Department of Computer Sciences, University of Wisconsin-Madison, WI, 53706*

## Abstract

Various machine learning models have been used to predict the properties of polycrystalline materials, but none of them directly consider the physical interactions among neighboring grains despite such microscopic interactions critically determining macroscopic material properties. Here, we develop a graph neural network (GNN) model for obtaining an embedding of polycrystalline microstructure which incorporates not only the physical features of individual grains but also their interactions. The embedding is then linked to the target property using a feed-forward neural network. Using the magnetostriction of polycrystalline $Tb_{0.3}Dy_{0.7}Fe_2$ alloys as an example, we show that a single GNN model with fixed network architecture and hyperparameters allows for a low prediction error of ~10% over a group of remarkably different microstructures as well as quantifying the importance of each feature in each grain of a microstructure to its magnetostriction. Such microstructure-graph-based GNN model therefore enables an accurate and interpretable prediction of the properties of polycrystalline materials.

[†]yliang@cs.wisc.edu

[*]jhu238@wisc.edu

[‡]These authors contribute equally to this work.

**Introduction**

Polycrystalline materials are ubiquitously used in everyday life and industry. The properties of such materials are governed not only by the atomic lattice structure within each grain, but also their microstructures which typically refer to the size (nm-μm), shape, orientation, and adjacency relation of the grains. Here, we develop a graph neural network[1,2] based machine learning model which enables an accurate prediction of the property of polycrystalline microstructures and quantifying the relative importance of each feature in each grain to the predicted property.

The application of machine learning models to predicting the properties of materials microstructure usually becomes necessary in two scenarios. First, prediction from physics-based models is too slow due to the large size or high complexity of the material system. For example, brute-force simulation of polycrystalline samples having thousands of grains would be impractically slow. Second, developing an accurate physics-based predictive model is too challenging due to the insufficient understanding of the underlying physical processes, e.g., for materials with high compositional/structural complexity or under complex/extreme conditions.

Existing machine learning models for predicting the properties of polycrystalline microstructures can be classified into two main types: statistical-descriptor-based[3–6] and image-based[7–14]. Both models focus on extracting statistical correlations between microstructures and properties, thereby achieving a fast prediction of the properties of similar microstructures via regression. A critical step in both models is to obtain a low-dimensional representation (i.e., embedding) of the microstructure, because raw microstructure data are high-dimensional and usually cannot directly be linked to the targe properties. In the statistical-descriptor-based machine learning models, microstructure is represented by statistical correlation function, such as a two-point correlation function[3–5] which relates two physical features (composition, crystal orientation, etc.) at two different spatial locations through a probability function[15]. In the image-based machine learning models, raw microstructure data can be directly used as the model input. For example, each voxel of a 3D microstructure image can be associated with a vector that stores the physical features (e.g., crystal orientation[12]) in that voxel. Convolutional neural network (CNN) can then be used to obtain low-dimensional microstructure embeddings (known as 'feature maps') in an automated manner with little bias from human researchers[7–14]. In both types of models, although the physical features at different positions/voxels of a microstructure are somewhat correlated, there are no reported means to let the physical features from neighboring grains interact with each other, because the adjacency relations of the grains are not stored. The inability to directly consider such interactions could negatively affect the performance of the subsequent property prediction, because the macroscopic properties of polycrystalline materials are critically determined by such microscopic interactions[16,17].

**Results**

**Building a Microstructure Graph**

In this article, we present a third type of machine learning model for predicting the properties of polycrystalline materials, where a polycrystalline microstructure is represented using a graph, which refers to a data structure comprising a set of interacting nodes. Figure 1 illustrates the conversion of a microstructure input into a graph input $G$ using a simple ten-grain microstructure as an example. As shown in Fig. 1, each grain is labelled and represented by a node. Physical features of each grain are stored using a vector, which together form a feature matrix $\mathbf{F}$, as sketched on the bottom right. There are five components in each feature vector, including three Euler angles ($\alpha$, $\beta$, $\gamma$) for describing the grain orientation, grain size which is defined as the number of voxels occupied by a specific grain, and the number of neighboring grains, as shown in the left panel of Fig. 1. These three physical features are selected because the physical properties of polycrystalline materials are critically determined by grain size, grain orientation, and the interactions among neighboring grains. If the grains have strong shape anisotropy (e.g., columnar grains), one may also include additional components in the feature vector such as the aspect ratio and orientations of the long and short axes of a grain, and the number of faces of the grain. All these physical features of grains can be obtained experimentally by for example the high-energy x-ray diffraction microscopy, which permits reconstruction of 3D crystallographic orientation fields from diffraction data and thereby extracting the size, 3D morphology, and Euler angles (or other parameters describing the orientations) of thousands of grain simultaneously[18,19].

The adjacent relation of all the nodes/grains is stored using a separate matrix $\mathbf{A}$, where matrix element $A_{ij}=1$ if grain $i$ and $j$ are neighbors (in physical contact) and $A_{ij}=0$ otherwise. In Fig. 1, the adjacency relation is illustrated by connecting neighboring grains/nodes with solid lines, as sketched on the top right. The feature matrix $\mathbf{F}$ and the adjacency matrix $\mathbf{A}$ together constitute the graph input of a microstructure, $G=(\mathbf{F}, \mathbf{A})$. By relating $\mathbf{F}$ and $\mathbf{A}$ using an appropriate mathematical function (known as the update function), physical features from neighboring grains can be correlated, as will be discussed later.

Based on such microstructure graph $G$, we develop a graph neural network (GNN)[1,2] model for predicting the microstructure-property link in polycrystalline materials. Using polycrystalline $Tb_{0.3}Dy_{0.7}Fe_2$ (Terfenol-D) alloys as an example, we train the GNN model using datasets of different sizes. Each dataset contains a group of different polycrystalline microstructures with corresponding magnetostriction (that is, macroscopic mechanical deformation induced by an applied magnetic field)[20]. A low prediction error of ~10% is achieved over a broad range of dataset, notably with relatively small datasets (down to only 72 microstructures). Relative importance of each feature of each grain in a given microstructure to the predicted magnetostriction is also quantified.

**Dataset Generation**

Our main goal is to demonstrate that one single GNN model with fixed network architecture and hyperparameters can be used to accurately predict the properties of a diverse set of polycrystalline microstructures. To this end, we use Dream.3D to generate 492 different 3D polycrystalline microstructures. The number of grains per microstructure varies broadly from 12 to 297. Microstructures with and without strong textures are both generated. Figure 2a-c show the statistical distributions of the size, the number of neighboring grains, and the orientation of all 87981 grains in the 492 microstructures. The grain orientation is represented by Euler angles $\alpha$, $\beta$, $\gamma$ in '$zxz$' sequence. For each microstructure (see an example in Fig. 2d), we performed phase-field modeling (see Methods) to obtain 3D distributions of local magnetization under a magnetic field applied along the $x$-axis (denoted as $H_x$). 2D slice of the local magnetization distribution under zero magnetic field is shown in Fig. 2e. Such magnetization distribution is determined by both the magnetic and elastic properties (elastic modulus, orientation) of individual grains as well as the magnetic and elastic interactions among both the neighboring and non-neighboring grains. Figure 2f shows the distribution of local magnetostriction $\lambda_{xx}$ associated with the magnetization distribution. The effective magnetostriction $\lambda_{xx}^{\text{eff}}$, which describes the macroscopic deformation of the material along the $x$-axis induced by the applied magnetic field, can then be obtained by taking volumetric average of the local $\lambda_{xx}$. For each microstructure, we set $\lambda_{xx}^{\text{eff}}=0$ at zero magnetic field ($H_x=0$) as a reference. As $H_x$ increases, $\lambda_{xx}^{\text{eff}}$ also increases (Fig.

2g) because magnetization vectors inside each grain will rotate to the direction of magnetic field. Once all magnetization vectors are parallel to the magnetic field, $\lambda_{xx}^{\text{eff}}$ would reach its saturation value. Four or five different magnetic fields are applied to each microstructure, amounting to 2287 data points. Each data point has a format of $[(G, H_x), \lambda_{xx}^{\text{eff}}]$, where $G$ refers to the graph input of the polycrystalline microstructure, which can be converted from the raw microstructure input as discussed above.

**Graph Neural Network built upon Microstructure Graph**

Since being introduced in 2009[21], over a dozen GNN models of different network architectures and/or update functions have been proposed for different application scenarios[1]. For materials-related research, GNN models have been developed for predicting the properties of molecules[22–26] and crystals[23,27], but not yet for polycrystalline materials to our knowledge. In this work, we employ a relatively simple but general GNN model, namely, the graph convolutional network (GCN)[28], to obtain the embedding of the polycrystalline microstructure graph $G$. This low-dimensional embedding incorporates not only the physical features of the grains but also their interactions. A fully connected layer is then used to link the embedding to the target property.

As shown in Fig. 3, the GCN model comprises a series of message passing layers (MPLs). Each MPL employs an update function, which updates the feature vector of each node based on its own feature vector and the feature vectors of its neighboring nodes. After the $n$th MPL, the input feature matrix of the graph $\mathbf{F}^{(n)}$ is updated to $\mathbf{F}^{(n+1)}$. The adjacency matrix $\mathbf{A}$ remains unchanged in the entire process. The layer-wise update function is given as[28],

$$\mathbf{F}^{(n+1)} = \sigma\left(\widehat{\mathbf{D}}^{-\frac{1}{2}}\widehat{\mathbf{A}}\widehat{\mathbf{D}}^{-\frac{1}{2}}\mathbf{F}^{(n)}\mathbf{W}^{(n)}\right). \tag{1}$$

Here $\sigma(x)=\max(0,x)$ denotes an activation function known as Rectified Linear Unit (ReLU), which is applied for every element x of the matrix in the bracket. $\hat{\mathbf{A}} = \mathbf{A} + \mathbf{I}$ is a summation of adjacency matrix A and an identity matrix $\mathbf{I}$. $\hat{\mathbf{D}}$ is the diagonal matrix with $\hat{D}_{ii} = \sum_j \hat{A}_{ij}$. The matrix $\hat{\mathbf{A}}$ undergoes a symmetric normalization $\hat{\mathbf{D}}^{-\frac{1}{2}}\hat{\mathbf{A}}\hat{\mathbf{D}}^{-\frac{1}{2}}$ before being multiplied with the input feature matrix $\mathbf{F}^{(n)}$. $\mathbf{W}^{(n)}$ is a trainable weight matrix. Depending on the dimension of $\mathbf{W}^{(n)}$, the dimension of the feature matrix may change after passing through the MPL. For example, after the second MPL in Fig. 3, the 10×5 matrix $\mathbf{F}^{(1)}$ transforms into a 10×7 matrix $\mathbf{F}^{(2)}$ after being multiplied with a 5×7 weight matrix $\mathbf{W}^{(1)}$. Overall, the matrix operation in the bracket of Eq. (1) enables taking a weighted average of the feature vectors of each node and those of its neighboring nodes. As a result, the interactions between the physical features of neighboring grains are considered in the output feature matrix $\mathbf{F}^{(n+1)}$.

The embedding of the graph input, denoted as {$\mathbf{X}$} in Fig. 3, can then be obtained by concatenating the most updated feature matrix $\mathbf{F}^{(n+1)}$. The {$\mathbf{X}$}, together with $H_x$, are used as the inputs of a fully connected layer (FL) for regressing the output property or properties {$\mathbf{Y}$} – in this case the effective magnetostriction $\lambda_{xx}^{\text{eff}}$. As shown in Fig. 3, each unit $v_i$ in the hidden layer or the output layer is determined by a weighted summation of all units $u_j$ in its preceding layer,

$$v_i = \sigma\left(\sum_j w_{ij}u_j + b_i\right), \tag{2}$$

where the activation function $\sigma(x)$ is again the ReLU. $w_{ij}$ is the weight and $b_i$ is the bias parameter, both of which can be trained to improve prediction performance.

**Property Prediction by the GNN model**

The whole dataset of the generated 2287 data points of $[(G, H_x), \lambda_{xx}^{\text{eff}}]$, in which there are 492 different graphs $G$ (Four or five $H_x$ are associated with each graph), are divided into 10 subsets of roughly the same size, including 8 subsets (1831 data points) for training, 1 subset (228 data points) for validation, and 1 subset (228 data points) for testing. Figure 4a shows the GNN-predicted $\lambda_{xx}^{\text{eff}}$ versus the true $\lambda_{xx}^{\text{eff}}$ of training and validation datasets. The hyperparameters are optimized to ensure the associated GNN model delivers the lowest macro average relative error (MARE) on the validation dataset. Details of model training and hyperparameters optimization are provided in the Methods section. Figure 4b shows the GNN-predicted $\lambda_{xx}^{\text{eff}}$ versus the true $\lambda_{xx}^{\text{eff}}$ of the independent testing dataset. The MARE for these data is as low as 8.05%. A data ablation test is performed to investigate the influence of dataset size on the MARE for the testing data points. A smaller group of microstructure graphs $G$ are randomly selected from the total 492 different graphs. The associated data points of $[(G, H_x), \lambda_{xx}^{\text{eff}}]$ are then used to evaluate the performance of the GNN model. As shown in Fig. 4c, the average value of the MARE quickly decreases from 17% to ~10% when the number of microstructure graphs increases to 72. As the dataset size further increases, the prediction error decreases more slowly and saturates at ~9%. This proves that the model is capable of achieving low prediction error with relatively small dataset. Moreover, the standard deviation of the MARE, represented by the error bar, decreases and then remains stable with increasing dataset size, indicating an improved stability of prediction.

**Computational efficiency of the GNN model**

It is generally more computationally efficient to use a trained GNN model for property prediction than using either physics-based direct numerical simulations such as phase-field modeling or image-voxel-based machine learning models such as CNN. In this work, predicting 1 data point of effective magnetostriction of one microstructure by phase-field modeling will take ~5 hours to complete with 16 cores running simultaneously on state-of-the-art supercomputers. By contrast, the property prediction by a trained GNN model is almost instantaneous: for example, it only takes ~0.2 s to predict all the 228 data points in Fig. 4b.

For evaluating the computational efficiency of our GNN model with respect to CNN models, three different 3D CNN models that have previously been utilized to predict the properties of polycrystalline[12] or two-phase[8,29] composite microstructures were trained separately under the same hardware (one Tesla P100 GPU core). Details of datasets generation for such tests, time measurement, and the CNN models are in the Methods section. Figure 5a compares the training time (the data loading time plus model training time) of our GNN model and the three CNN models as a function of the number of voxels in the 3D microstructure $N_d{}^3$. In CNN models, the representation of microstructure is a voxel-by-voxel basis. Therefore, the total training time increases significantly with the total number of voxels. Such increase is particularly significant for CNN models with complex network architecture and a large number of convolutional or pooling layers such as the pre-trained VGGNet (denoted as the CNN-3 in Fig. 5a). In the meantime, the cache memory required by CNN models also increases significantly with the number of voxels. Specifically, the cache memory of a single Tesla P100 GPU core would be insufficient for training the CNN-1, CNN-2, and CNN-3 model when the number of voxels exceed $112^3$, $120^3$, $64^3$, respectively. This severely limits the capability of applying 3D CNN models to large-scale microstructure image datasets, which often have 1 billion (=$1000^3$) voxels or more per image[19].

By comparison, microstructure representation is a node-by-node basis in GNN models. Therefore, the training time is directly determined by the number of nodes in a graph, which is the number of grains ($N_{grain}$) in a polycrystalline microstructure. In every 3D microstructure image generated for computational efficiency tests in Fig. 5a, $N_{grain}$ is fixed to be 300 although the number of voxels ($N_d^3$) can be different. As a result, the total training time of our GNN model does not vary with the number of voxels. In particular, since a 3D polycrystalline microstructure image almost always contains a greater number of voxels than the number of grains (that is, one grain occupies more than one voxel), the total training time of CNN should typically be longer than the GNN. In the present test (Fig. 5a), the training time of CNN-2 model is about 35 times longer than that of GNN for a 3D microstructure image containing $120^3$ voxels and 300 grains, and this difference in training time is expected to further increase in larger-scale microstructure images. Figure 5b shows the training time of our GNN model as a function of the $N_{grain}$, with the number of features in each grain remaining unchanged. As shown in Fig. 5b, our GNN model can be trained with a large-scale polycrystalline microstructure containing up to 4700 grains without exceeding the cache memory limit of one single Tesla P100 GPU, with an acceptable total training time of ~13 hours for a training dataset of 160 data points.

**Discussion**

With a trained GNN model, the importance of each feature in each node of a graph to the predicted output can be quantified using the Integrated Gradient (IG) method[30]. Specifically, one can evaluate the IG value for each physical feature in each node/grain of a microstructure graph $G$. The higher the absolute IG value, the more important the specific feature. As illustrated in Fig. 6a, the IG value of a given feature (grain size in this case) is calculated by integrating the gradient of the predicted magnetostriction with respect to the feature along a straight line from a baseline graph $\tilde{G} = (\tilde{\mathbf{F}}, \mathbf{A})$ to the target graph $G = (\mathbf{F}, \mathbf{A})$; see details in Methods section.

Among the three physical features of a grain (size, orientation, the number of neighboring grains) being considered in the present GNN model, the grain size and the number of neighbors are somewhat correlated in a computer-generated microstructure with fixed dimension: in general, the bigger the grain size, the more neighboring grains. Therefore, we focus on comparing the importance of grain size and grain orientation to the effective magnetostriction through the IG analysis. However, since the grain orientation is described by three different feature components (the Euler angles $\alpha$, $\beta$, $\gamma$), one cannot use the IG value of one specific Euler angle to represent the importance of grain orientation. Simply averaging the IG values of all three Euler angles would not be physically reasonable as well.

To address this issue, a two-step analysis is performed. The first step is to calculate the IG values of the grain size with a specific baseline graph $\tilde{G}$, in which the grain size is set to be zero in all nodes (see Fig. 6a) while the values of the other four node features ($\alpha$, $\beta$, $\gamma$, and number of neighbors) are set to be the same as those of the target graph $G$. Using such a baseline graph allows us to quantify the individual contribution of grain size to the macroscopic property. As an example, Figure 6b shows the absolute IG value of each feature of each node/grain in a simple microstructure graph with 12 nodes/grains, where only the grain size (the second to the last in the five-component node feature vector) displays nonzero absolute IG values. The second step is to analyze the correlation between the grain size and the absolute IG value of grain size for each microstructure graph. Figure 6c shows such correlation analysis in the abovementioned 12-node microstructure graph. A relatively strong correlation is observed, with a $R^2$ factor of >0.9 in both linear and polynomial regression. This indicates that the grain size is more important than grain orientation in this specific microstructure graph. This is due to the fact that increasing the absolute value of normalized grain size would always enhance the absolute IG value of grain size if omitting the contribution from the grain orientation, as can be seen from the Eq. (10) in the Methods section. In this regard, if there is no strong correlation, the only possible explanation would be that the contribution of grain orientation outweighs the contribution of grain size. Figure 6d shows the distribution of the $R^2$ factor among the 492 microstructure graphs. It can be seen that the majority of them only display a relatively weak correlation. This means that in the present dataset, grain orientation generally plays a more important role than grain size in determining the effective magnetostriction. This is consistent with the physical principle that the magnetostriction would be different in grains of same grain size but different orientations.

In summary, a graph neural network (GNN) model has been developed for predicting the effective properties of polycrystalline materials, where a polycrystalline microstructure is represented by a graph. Using the effective magnetostriction of Terfenol-D alloys as an example, the GNN model demonstrates a low relative error of 8.05% in property prediction with a dataset of ~500 polycrystalline microstructures in which the number of grains per microstructure vary from 12 to 297. Therefore, a single GNN model with fixed network architecture and hyperparameters can be used to rapidly and accurately predict the magnetostriction of these ~500 different microstructures. In particular, the prediction error of this GNN model can remain at a low level of ~10% over a broad range of dataset size, even when the dataset size is relatively small. In combination with the Integrated Gradient (IG) method, our GNN model enables quantifying the importance of each feature in each grain to the predicted effective properties, which has remained challenging to achieve with existing data-driven or physics-based models. Importantly, the design of baseline graph for the calculation of IG values and the post-processing of the calculated IG values need to align with the physical principles of the target material system.

Overall, we show that graph-based representation of a polycrystalline microstructure permits a direct consideration of the physical interactions among neighboring grains, which leads to an accurate property prediction by the associated GNN model. Moreover, the grain-by-grain analysis of the microstructure in the present GNN model is computationally more efficient than the voxel-by-voxel analysis of microstructure in existing machine learning models such as CNN, because one grain often occupies multiple voxels in a reasonably high-spatial-resolution microstructure image. This high computational efficiency will become particularly valuable for harnessing large-scale 3D polycrystalline microstructures, e.g., those obtained through high-energy x-ray diffraction microscopy[19], which typically contain billions of voxels per microstructure image. This unique combination of high accuracy, high interpretability, and high computational efficiency makes our microstructure-graph-based GNN model an attractive tool for harnessing large-scale 3D microstructure datasets.

Looking ahead, the present GNN model can be extended to describe the orientations (including 3 parameters for the crystal misorientation and 2 parameters for the orientation of the normal axis) and other features of grain boundaries[31] by assigning an additional feature vector $e_{ij}$ to each edge that connects node (grain) $i$ and node (grain) $j$ in the microstructure graph. This is similar to the application of GNN model for predicting the properties of organic molecules and inorganic crystals[23,27] where an edge vector $e_{ij}$ is introduced to describe the features of the chemical bond between atom $i$ and atom $j$. The present GNN model can also be extended to describe the features of other mesoscale defects (e.g., precipitates and pores) by introducing additional nodes in the microstructure graph such that the defect-grain interactions can be incorporated. Therefore, graph can enable a compact but comprehensive representation of the salient features of the hierarchical structure elements in a polycrystalline microstructure. However, for the subsequent development of GNN model on effective property prediction, the appropriate level of complexity in the microstructure graph, the selection of physical features for the grains and/or grain boundaries, the identification of optimum network architecture and update functions will critically depend on the physical principles of the target material systems or properties. It is challenging to determine one GNN model with fixed microstructure graph input, network architecture, and update function (the three key components of a GNN model) that is universally applicable to any polycrystalline materials and/or effective properties. In fact, developing such universal machine learning models for harnessing vastly different datasets remains to be one of the most significant challenges in the entire field of artificial intelligence. Despite this, when the available training datasets for the target material system is not abundant, it would be beneficial to use a GNN model pre-trained using larger datasets of similar materials/properties as the starting point of model development. Such transfer learning of GNN models for predicting the properties of polycrystalline microstructures has not yet been reported, but can be pursued by leveraging the established strategies of pre-training GNN models for molecular property prediction[32]. We thus hope the present work can stimulate more research into the development and application of graph-based machine learning models for predicting the microstructure-property link and other important tasks (e.g., predicting processing-microstructure link and microstructure evolution) in the field of microstructure informatics[33].

**Methods**

**Phase-field simulations of the effective magnetostriction of polycrystals**

Phase-field simulations are performed using the commercial μ-Pro® package (mupro.co). In the simulation, the whole polycrystalline microstructure is discretized into $64 \times 64 \times 64$ cuboid-shaped cells with dimension of $1\,\text{nm} \times 1\,\text{nm} \times 1\,\text{nm}$. In each cell, the magnetization in the global system (sample coordinate system) $\mathbf{M} = M_s\mathbf{m}$, where $M_s$ is saturation magnetization and $\mathbf{m} = (m_x,\ m_y,\ m_z)$ is the normalized magnetization vector. The total free energy of the polycrystal $F_{\text{tot}}$ is written as

$$F_{\text{tot}} = \int_V \left( f_{\text{exch}} + f_{\text{stray}} + f_{\text{ext}} + f_{\text{anis}} + f_{\text{elastic}} \right) dV, \tag{3}$$

where $f_{\text{exch}}$, $f_{\text{stray}}$, $f_{\text{ext}}$, $f_{\text{anis}}$ and $f_{\text{elastic}}$ are the densities of Heisenberg exchange energy, magnetic stray field energy, external field energy, magnetic anisotropy energy and elastic energy, respectively. Among them, $f_{\text{exch}} = A(\nabla\mathbf{m})^2$, where $A$ is the Heisenberg exchange constant. $f_{\text{stray}} = -\frac{1}{2}\mu_0 M_s \mathbf{H}_{\text{stray}} \cdot \mathbf{m}$, where $\mu_0$ is the vacuum permeability and $\mathbf{H}_{\text{stray}}$ is the magnetic stray field. Periodic boundary condition is used in the simulation, hence $\mathbf{H}_{\text{stray}} = -\nabla\phi + \mathbf{N}_{\text{D}}\overline{\mathbf{M}}$, where $\phi$ is the magnetic scalar potential solved using the Fourier spectral method[34], $\mathbf{N}_{\text{D}}$ is the demagnetizing factor which depends on the shape of the microstructure and $\overline{\mathbf{M}}$ is the average magnetization of the whole microstructure. External field energy density $f_{\text{ext}} = -\frac{1}{2}\mu_0 M_s \mathbf{H}_{\text{ext}} \cdot \mathbf{m}$, where $\mathbf{H}_{\text{ext}}$ is the external magnetic field.

To calculate $f_{\text{anis}}$ and $f_{\text{elastic}}$ in global system, the normalized magnetization vector in local crystalline system $\mathbf{m}^L$ is introduced and can be converted from $\mathbf{m}$ through $m_i^L = R_{ij}m_j$ $(i, j = x, y, z)$, where $R_{ij}$ is the rotation matrix (determined by the Euler angles in each simulation cell) describing the difference between local crystalline system and global system[35].

Therefore, the cubic anisotropy energy density $f_{\text{anis}} = K_1 \left[ \left( m_x^L m_y^L \right)^2 + \left( m_y^L m_z^L \right)^2 + \left( m_z^L m_x^L \right)^2 \right] + K_2 \left( m_x^L m_y^L m_z^L \right)^2$, where $K_1$ and $K_2$ is the first-order and second-order magnetic anisotropy constant, respectively[34]. $f_{\text{elastic}} = \frac{1}{2} c_{ijkl} \left( \varepsilon_{ij} - \varepsilon_{ij}^0 \right) \left( \varepsilon_{kl} - \varepsilon_{kl}^0 \right)$, where $c_{ijkl}$ represents the elastic stiffness constant, $\varepsilon_{ij}$ is the total strain and $\varepsilon_{ij}^0$ is the spontaneous strain in the global system[36]. $\boldsymbol{\varepsilon}^0$ can be obtained from the spontaneous strain in local crystalline system $\boldsymbol{\varepsilon}^{0L}$ through $\varepsilon_{ij}^0 = R_{ki} R_{lj} \varepsilon_{kl}^{0L}$, where,

$$\varepsilon_{ij}^{0L} = \begin{cases} \frac{3}{2} \lambda_{100} \left( m_i^L m_j^L - \frac{1}{3} \right) & (i = j) \\ \frac{3}{2} \lambda_{111} m_i^L m_j^L & (i \neq j) \end{cases}. \tag{4}$$

$\lambda_{100}$ and $\lambda_{111}$ are the magnetostriction constants[34]. The total strain $\boldsymbol{\varepsilon}$ is the sum of spatially independent homogeneous strain $\bar{\boldsymbol{\varepsilon}}$ and a spatially heterogeneous strain $\boldsymbol{\varepsilon}^{\text{het}}$(ref. [37]). The distribution of $\boldsymbol{\varepsilon}^{\text{het}}$ at mechanical equilibrium can be obtained by numerically solving $\nabla \cdot \mathbf{c}(\boldsymbol{\varepsilon} - \boldsymbol{\varepsilon}^0) = 0$ through the Fourier Spectral Iterative Perturbation Method[35,38]. The homogeneous strain indicates the average deformation of the entire microstructure under an applied strain. In the simulation, no external strain is applied, hence, one can derive $\bar{\boldsymbol{\varepsilon}} = \bar{\boldsymbol{\varepsilon}}^0$ through minimizing the total elastic energy as a function of homogeneous strain. This also means, for a polycrystalline magnetic material that is stress-free in all directions, the local magnetostriction (magnetic-field-induced strain) $\lambda_{xx}$ is equal to the local spontaneous strain $\varepsilon_{xx}^0$, yielding,

$$\lambda_{xx} = \varepsilon_{xx}^0 = R_{kx} R_{lx} \varepsilon_{kl}^{0L}, \qquad \lambda_{xx}^{\text{eff}} = \bar{\varepsilon}_{xx}^0 \tag{5}$$

Therefore, for obtaining the $\lambda_{xx}$ and $\lambda_{xx}^{\text{eff}}$, one needs to obtain the equilibrium distribution of local magnetization $\mathbf{m}^L$ under different magnetic fields. To this end, we compute the evolution of the local magnetization $\mathbf{m}$ (in global system) induced by effective magnetic field $\mathbf{H}_{\text{eff}} = -\frac{1}{\mu_0 M_s}\frac{\delta F_{\text{tot}}}{\partial \mathbf{m}}$ by solving the Landau-Lifshitz-Gilbert (LLG) Equation,

$$\frac{\partial \mathbf{m}}{\partial t} = -\frac{\gamma_0}{1+\alpha^2}(\mathbf{m}\times \mathbf{H}_{\text{eff}} + \alpha\,\mathbf{m}\times\mathbf{m}\times \mathbf{H}_{\text{eff}}) \qquad (6)$$

where $\alpha$ is the Gilbert damping coefficient and $\gamma_0$ is the gyromagnetic ratio. The LLG equation is solved using implicit Gauss–Seidel projection method implemented with Fourier-Spectral approach[34], with a dimensionless discretized time interval $\Delta t^* = \frac{\gamma_0 M_s}{1+\alpha^2}\Delta t = 0.005$. The material parameters of the Terfenol-D used for phase-field simulations are listed as follows[34]: saturation magnetization $M_s = 8.0 \times 10^5$ A m$^{-1}$, exchange constant $A = 9.0 \times 10^{-12}$ J m$^{-1}$, anisotropy constant $K_1 = -6.0 \times 10^4$ J m$^{-3}$, $K_2 = 0$; elastic stiffness constant $c_{11} = 1.41 \times 10^{11}$ Pa, $c_{12} = 6.48 \times 10^{10}$ Pa, $c_{44} = 4.87 \times 10^{10}$ Pa, magnetostriction constant $\lambda_{100} = 0$, $\lambda_{111} = 1.64 \times 10^{-3}$. A randomized magnetization distribution is considered as the initiate state. The equilibrium distribution of $\mathbf{m}$ is considered to be reached when the total free energy density $f_{\text{tot}}$ no longer changes significantly with time.

**GNN model training and evaluation protocols**

All trainable weights in the model are updated through Gradient Descent,

$$w = w - \eta\frac{\partial L}{\partial w} \qquad (7)$$

where $w$ includes $\mathbf{W}^{(n)}$ in MPLs (Eq. 1); the $w_{ij}$ and $b_i$ in the fully connected layers (Eq. 2); $\eta$ is the learning rate, $L$ is the loss, taken as the mean square error (MSE) between the predicted value $\hat{y}_m$ and real value $y_m$ in one batch,

$$L = \frac{1}{n} \sum_{m=1}^{n} (\hat{y}_m - y_m)^2, \qquad (8)$$

where $n$ is the batch size – the number of data points used for one iteration of weight updating. During the training process, the whole training dataset will be iterated several times. The number of iterations is called the number of epochs. The gradient $\partial L / \partial w$ is obtained through back propagation. The macro average relative error (MARE) is used for representing the relative error between the model predicted value and the simulated value of testing dataset (see Fig. 4c),

$$\text{MARE} = \frac{\sum_{m=1}^{n_0} |\hat{y}_m - y_m|}{\sum_{m=1}^{n_0} |y_m|}, \qquad (9)$$

where $n_0$ is the total number of testing data points.

Hyperparameter optimization is performed with the whole dataset with 492 microstructures. The associated data points of the form $[(G, H_x), \lambda_{xx}^{\text{eff}}]$ are randomly divided into 10 subsets of about the same size. Among them, 1 subset is used for validation, 1 subset is used for testing and the remaining 8 subsets are used for training. 81 different GNN models, each with a unique set of hyperparameters (including the dimension of the weight matrices $\mathbf{W}^{(1)}$ and $\mathbf{W}^{(2)}$ in MPLs, the number of units in the first hidden layer $N_{h1}$ and the second hidden layer $N_{h2}$ in the FL, batch size $n$, number of epochs, and learning rate $\eta$), are trained. These 81 trained GNN models are then independently evaluated for MARE with the validation dataset. The GNN model that leads to the lowest MARE is identified as the optimized model and applied to the testing dataset, which has remained unseen during the entire training and validation process, for the evaluation of property prediction performance (indicated by the 'Testing MARE' in Fig. 4b). The hyperparameters associated with this optimized GNN model are listed in Table 1.

For the data ablation study in Fig. 4c, the optimized hyperparameters listed in Table 1 were used for training the GNN model in each case. 10-fold cross validation is used to evaluate the prediction performance. A group of $k$ microstructure graphs ($k$ varying from 32 to 392) are randomly selected from the 492 microstructure graphs. For each group, the associated data points (ca., 4$k$-5$k$) of the form [($G$, $H_x$), $\lambda_{xx}^{\text{eff}}$] are randomly divided into 10 subsets of about the same size. Each time, 1 subset is used for testing while the remaining 9 subsets are used for training. Such training and testing processes are repeated for 10 times, and it is ensured that each subset is only used for testing for one time. The resultant 10 MSE and MARE numbers are averaged to be the final estimation.

Finally, to evaluate the stability of the optimized GNN model, the influence of randomness on the model performance is studied, including (1) the random selection of the group of $k$ microstructures graphs from the total 492 graphs; and (2) the random division of the associated data points (ca., 4$k$-5$k$) into 10 subsets. Specifically, starting from the GNN models with optimized hyperparameters (Table 1), we used 3 different random numbers to achieve three different versions of the 10 subsets and then performed 10-fold validation independently. The average and standard deviation of MARE are calculated based on the obtained 30 testing errors (Fig. 4c).

**Time measurement of the GNN models and CNN models**

The three CNN models used for time measurement (in Fig. 5a) have the same neural network architecture as those in the literature: Ref. [8] for CNN-1; Ref. [29] for CNN-2; Ref. [12] for CNN-3. The inputs to these three CNN models are 3D microstructure images. Each image is comprised of 3D discretized voxels of $N_d \times N_d \times N_d$, where each voxel is assigned to three values of Euler angles. The whole dataset for time measurement comprises 50 different 3D microstructure images. Together with the 4 different magnetic fields $H_x$ applied to each microstructure and the associated effective magnetostriction $\lambda_{xx}^{\text{eff}}$, a total of 200 data points is used. Each point has a format of [($I_{3D}$, $H_x$), $\lambda_{xx}^{\text{eff}}$]. The whole dataset is then divided into 8 training subsets (160 data points), 1 validation subset (20 data points), and 1 testing subset (20 data points). For a fair comparison of the model training time, the batch size and number of epochs in the three CNN models and the GNN models are all set to be the same as those listed in Table 1. All tasks (data loading and model training) were performed on a single Tesla P100 GPU core. We note that the CNN-2, due to its specific network architecture, are not applicable to microstructure images of some $N_d$ values (specifically, $N_d$=32), hence its data loading and training time cannot be measured.

**Integrated Gradient (IG) calculation**

The recently proposed Integrated Gradient (IG) method[30] is used to evaluate the importance of each physical feature in each grain to the predicted magnetostriction for a given microstructure graph $G$=($\mathbf{F}$, $\mathbf{A}$). The microstructure graph $G$ along with the external field $H_x$ is termed as the real input ($G$, $H_x$) below. The calculation of the IG values for each feature requires defining a baseline input ($\tilde{G}$, $H_x$), where $\tilde{G}$= ($\tilde{\mathbf{F}}$, $\mathbf{A}$), which is analogues to the selection of the origin of a coordinate system. Specifically, the grain size in the feature matrix $\tilde{\mathbf{F}}$ of the baseline input are set to be zero while other features are set to the same with $\mathbf{F}$ of the real input. This allows us to quantify the individual contribution of grain size to the magnetostriction.

The integrated gradient (IG) of each feature in each grain is defined to be the aggregation of the gradients along the straight line connecting the baseline input with the real input, given as,

$$IG\left(f_i^k\right) ::= \left(f_i^k - \tilde{f}_i^k\right) \times \sum_{l=1}^m \frac{\partial y\left[\left(\tilde{\mathbf{F}} + \frac{l}{m} \times (\mathbf{F} - \tilde{\mathbf{F}}), \mathbf{A}\right), H_x\right]}{\partial f_i^k} \times \frac{1}{m} \qquad (10)$$

where $f_i^k$ and $\tilde{f}_i^k$ are the $k$th feature of $i$th node of graph of real input and baseline input,

respectively. Among the five features, the grain size ($k$=4) and number of neighbors ($k$=5) of each node in the microstructure graph are the normalized values of the original feature $f_{oi}^k$, specifically, $f_i^k = (f_{oi}^k - \mu)/\sigma$, where the average value $\mu$ and the standard deviation $\sigma$ are calculated based on the original feature values in all nodes of a specific microstructure graph.

The purpose of this normalization is to convert the two features into a common scale and meanwhile preserve the relative difference of these feature values across the nodes. $\mathbf{A}$ is the adjacency matrix of the real input; $y[(\tilde{\mathbf{F}}, \mathbf{A}), H_x]$ refers to the trained GNN model; $m$ is the number of steps in the Riemman approximation of the integral. As shown in Eq. (10), $IG\left(f_i^k\right)$ is directly determined by the $f_i^k$ through the term $\left(f_i^k - \tilde{f}_i^k\right)$ and also the aggregation term

$\sum_{l=1}^m \frac{\partial y\left[\left(\tilde{\mathbf{F}} + \frac{l}{m} \times (\mathbf{F} - \tilde{\mathbf{F}}), \mathbf{A}\right), H_x\right]}{\partial f_i^k} \times \frac{1}{m}$. The other features $f_{j \neq i}^k$ can indirectly influence the $IG\left(f_i^k\right)$ since

the gradient is influenced by all the features along the integration path. Note that the sum of all

IG values of the whole graph, that is, $\sum_{i=1}^N \sum_{k=1}^M IG\left(f_i^k\right)$, is equal to the difference between the two final predictions (the value of normalized $\lambda_{xx}^{\text{eff}}$ herein) obtained using the real input [($\mathbf{F}$,

$\mathbf{A}$), $H_x$] and the baseline input [$(\tilde{\mathbf{F}}, \mathbf{A}), H_x$]. Here $N$ denotes the number of nodes (grains) in the

graph (microstructure); $M$ is the number of features of each node/grain ($M$=5 herein).A large

positive (negative) IG value indicates that the predicted output will increase (decrease)

significantly as the value of a given feature increases. Therefore, larger absolute value of IG

suggests greater importance of a specific feature.

## Data Availability

The microstructure-property dataset of the 492 microstructures that support the findings of this study are available in GitHub via https://github.com/mehmetfdemirel/PolycrystalGraph. The other data that support the plots presented in this paper are available from the corresponding authors upon reasonable request.

## Code Availability

The open-source codes for the Graph Neural Network (GNN) model and Integrated Gradient (IG) calculations can be accessed via https://github.com/mehmetfdemirel/PolycrystalGraph.

## Acknowledgement

Acknowledgment is made to the donors of The American Chemical Society Petroleum Research Fund for partial support of this research, under the award PRF # 61594-DNI9 (M.D. and J.-M.H.). This work was supported in part by FA9550-18-1-0166, IIS-2008559, and NSF 1740707. J.-M.H. also acknowledges the support from a start-up grant from the University of Wisconsin-Madison. This work used the Extreme Science and Engineering Discovery Environment (XSEDE) Specifically, it used the Bridges-2 system at the Pittsburgh Supercomputing Center (PSC).

## Competing Interests

The authors declare no competing interests.

## Author Contributions

J.-M.H. initiated the project and designed the structure of the paper. J.-M.H and Y. L. supervised the research. M.D. and M.F.D. developed the codes for Graph Neural Network (GNN) model based on an existing version of the code from the research group of Y. L.. Specifically, M.F.D developed the code for the data input, GNN model architecture, model training and model testing. M.D. developed the code for data division and Integrated Gradient (IG) values calculation. M.D. generated the datasets and performed the phase-field simulations. Using substantial feedback from all coauthors, M.D. performed the GNN model training and testing, the IG analysis, and the benchmark test on computational efficiency against baseline convolutional neural network (CNN) models. M.D. and J.-M.H. wrote the paper with feedback from M.F.D. and Y. L..

**Figure Legends**

**Figure 1**. **Graph-based representation of a *N*-grain polycrystalline microstructure**. A 10-grain microstructure is considered as an example (*N*=10). The *M* different physical features (*M*=5 herein) of each grain (label '*i*', *i*=1,2,…*N*) are stored using *N* different *M*-component feature vectors $f_i^k$ (*k*=1,2…*M*), as shown in the left panel. These feature vectors are then combined into a *N*×*M* feature matrix **F**. The adjacency relations of the grains (nodes) are stored using an adjacency matrix **A**; the matrix element $A_{ij}$ equals 1 if grain *i* and grain *j* are neighbors and equals 0 otherwise. The neighboring grains are connected by lines (in top right). Overall, a graph *G*=(**F**, **A**).

**Figure 2. Dataset of 3D polycrystalline microstructures and their effective magnetostriction.** Distributions of (**a**) size (represented by the number of voxels), (**b**) number of neighboring grains for a specific grain, and (**c**) three Euler angles of the 87981 grains in the entire dataset of 492 microstructures. (**d**) An example of 3D polycrystalline microstructure in the dataset (colored by grain orientation). (**e**) Local magnetization and (**f**) local magnetostriction $\lambda_{xx}$ on the 2D slice shown in (**d**). (**g**) Effective magnetostriction $\lambda_{xx}^{\text{eff}}$ of the 3D microstructure shown in (**d**) as a function of the applied magnetic field $H_x$.

**Figure 3. Network architecture of the GNN model.** A series of message passing layers (MPLs) are employed for obtaining an embedding {**X**} of the graph input by sequentially updating the feature matrix (sketched in the bottom). The first MPL allows a specific node to take messages from its neighbors. By using more MPLs, interactions between non-neighboring nodes can also be established. For example, grain #1 and #10 are interacting through the intermediate grain #8. Such message passing enables considering the influence of all the other nodes in the graph on a specific node, which in effect describes the magnetostatic interaction among different grains. A simple graph input with 10 nodes (grains) is used for illustration. The embedding {**X**}, together with the applied magnetic field $H_x$, is linked to the effective magnetostriction $\lambda_{xx}^{\text{eff}}$ through a fully connected layer (FL).

**Figure 4. Property prediction by the trained GNN model.** (**a**) Predicted effective magnetostriction $\lambda_{xx}^{\text{eff}}$ vs. ground truth of training and validation datasets. (**b**) Predicted $\lambda_{xx}^{\text{eff}}$ vs. ground truth of the independent testing dataset. The GNN model used in (a-b) are trained,

validated and tested based on a dataset of 492 microstructures (2287 data points). (**c**) Macro Average Relative Error (MARE) of the GNN models trained using dataset of different size. Lower MARE value indicates better prediction performance. The error bar indicates the standard deviation of the MARE.

**Figure 5. Computational efficiency of the GNN model.** (a) The training time (the data loading time plus model training time) of our GNN model and the three baseline CNN models as a function of the number of voxels in the 3D polycrystalline microstructure, denoted as $N_d{}^3$, with $N_d$ varying from 32 to 128. The number of grains in each microstructure image, denoted as $N_{grain}$, is fixed to be 300. The training time is shown in reduced unit (r.u.), using the training time of the GNN model as the reference (training time=1). (b) The training time of our GNN model as a function of $N_{grain}$ in the range of 300 to 4700. The number of features in each grain/node is fixed to be 5. The training time is shown in real unit (seconds).

**Figure 6. Integrated Gradient (IG) analysis.** (**a**) schematic of the IG calculation by defining a baseline graph $\tilde{G}$, where the grain size is set to be zero in all nodes while the values of the other four node features are set to be the same as those of the target graph $G$. (**b**) Node-resolved absolute IG values of a 12-node microstructure graph. (**c**) Correlation analysis between the absolute value of the normalized grain size and its corresponding absolute IG value for this 12-node graph by both linear and polynomial fitting. (**d**) The distribution of the $R^2$ factor obtained through linear fitting and polynomial fitting for all the 492 microstructure graphs.

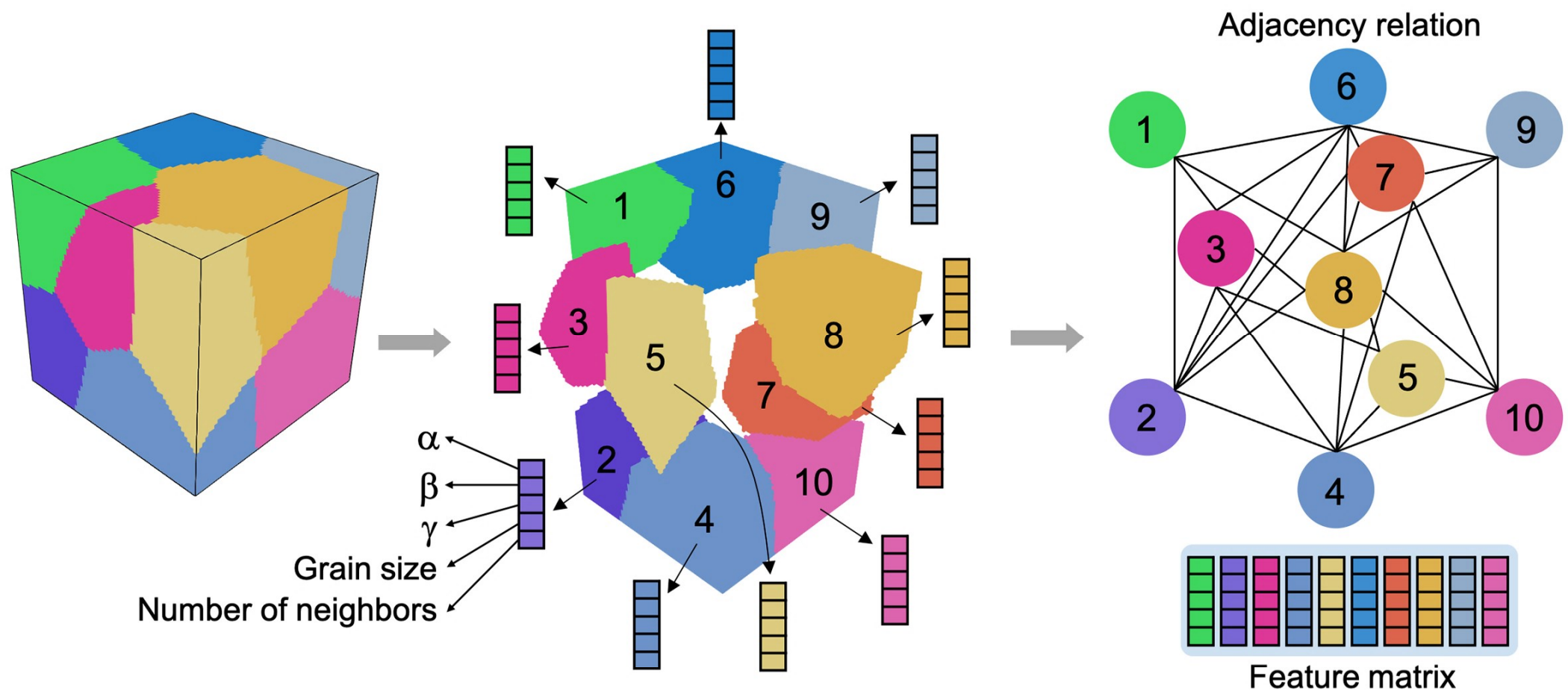

Adjacency relation

α
β
γ
Grain size
Number of neighbors

Feature matrix

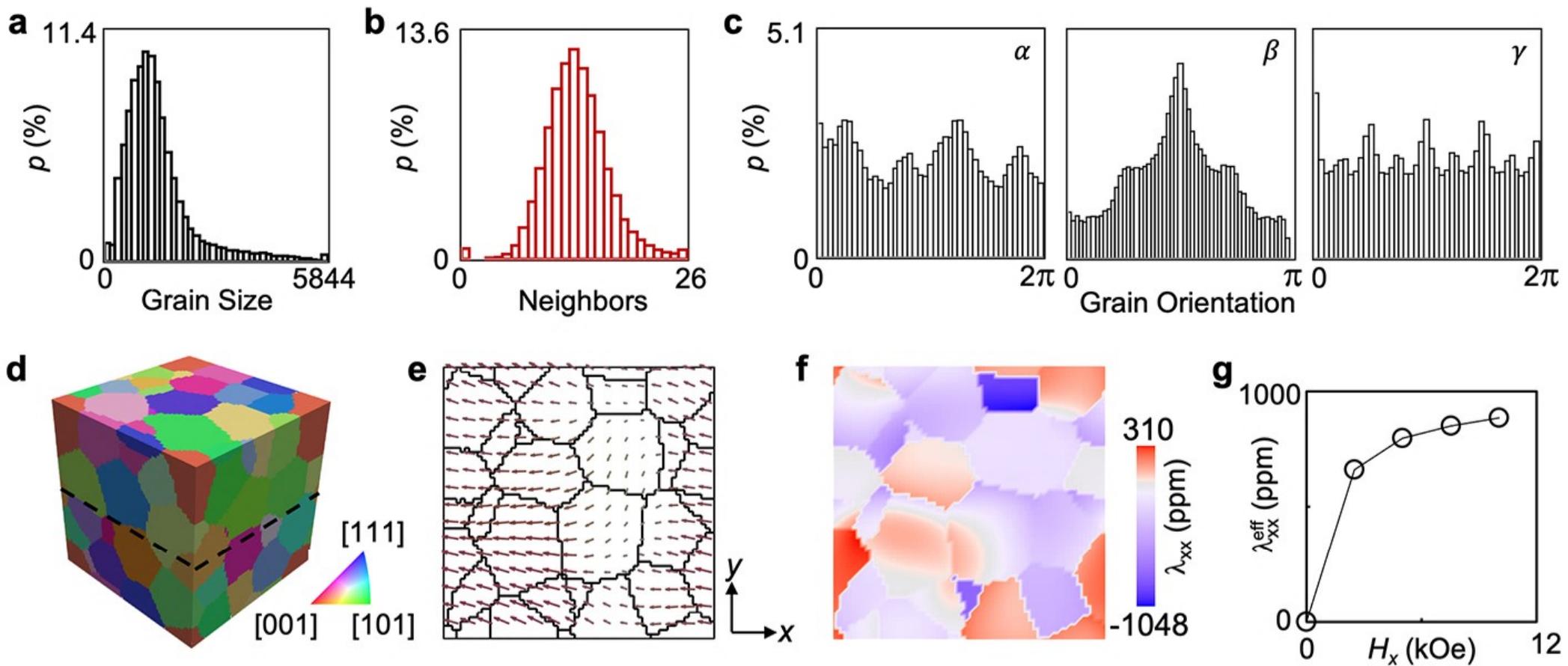

**a**

11.4

$p$ (%)

Grain Size

0　　　5844

**b**

13.6

$p$ (%)

Neighbors

0　　　26

**c**

5.1

$p$ (%)

$\alpha$　　$\beta$　　$\gamma$

0　　2π　0　　π　0　　2π

Grain Orientation

**d**

[111]

[001]　[101]

**e**

$y$

$x$

**f**

310

$\lambda_{xx}$ (ppm)

-1048

**g**

1000

$\lambda_{xx}^{\text{eff}}$ (ppm)

0

0　　$H_x$ (kOe)　12

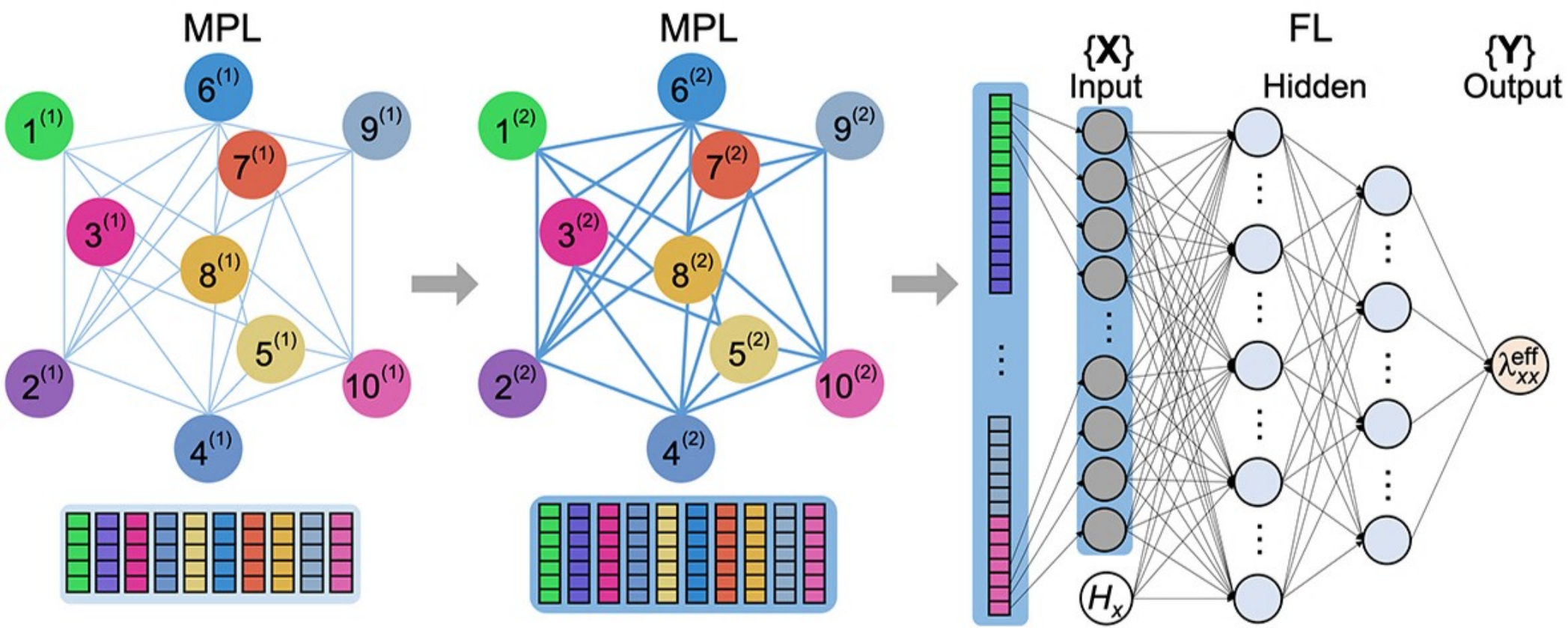

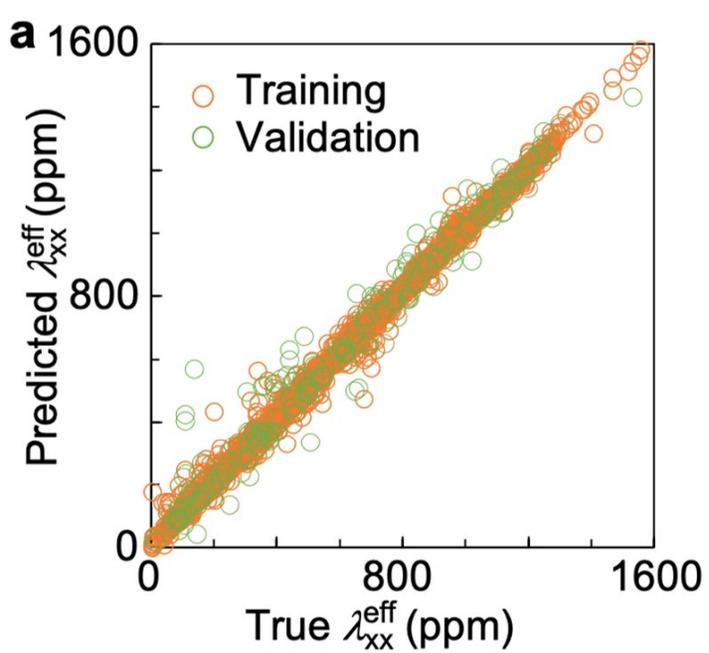
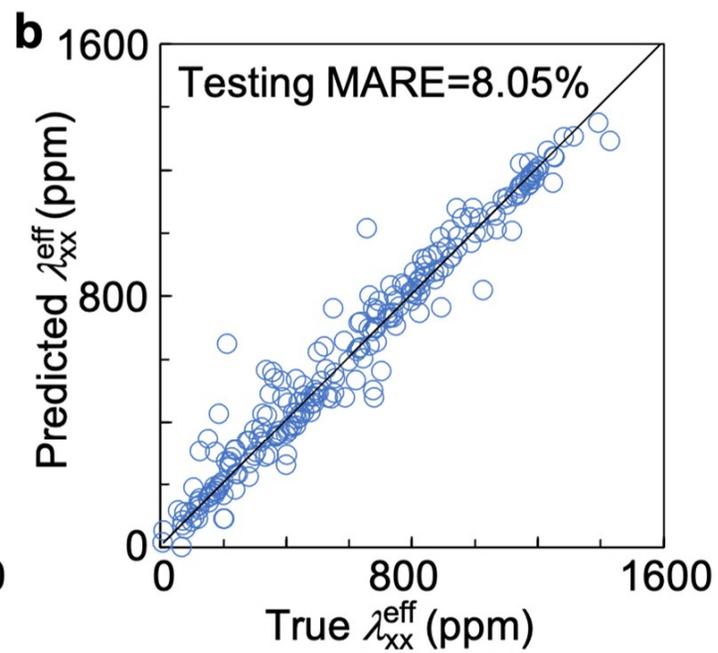
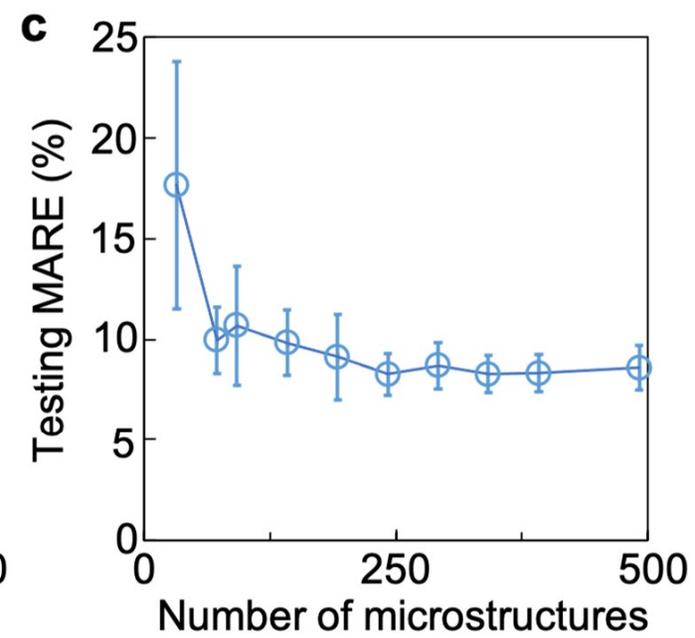

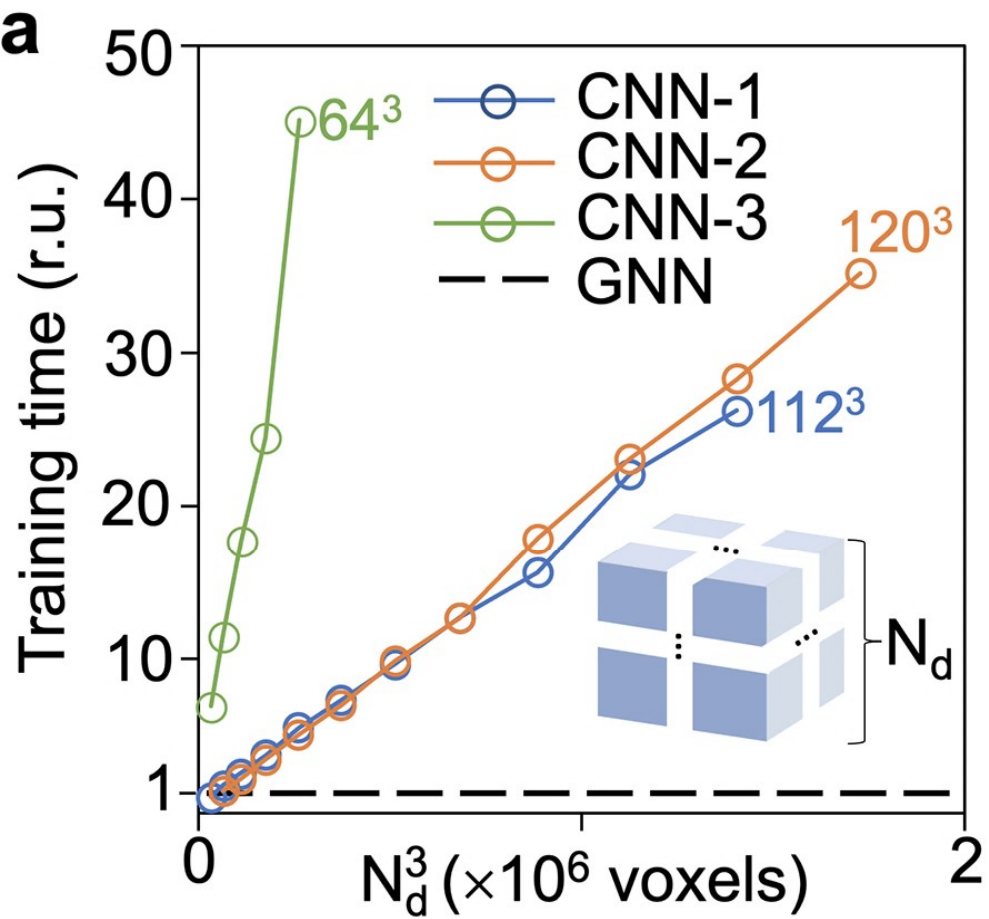

**a**

Training time (r.u.)

$N_d^3$ (×10⁶ voxels)

- CNN-1
- CNN-2
- CNN-3
- GNN

64³  120³  112³

$N_d$

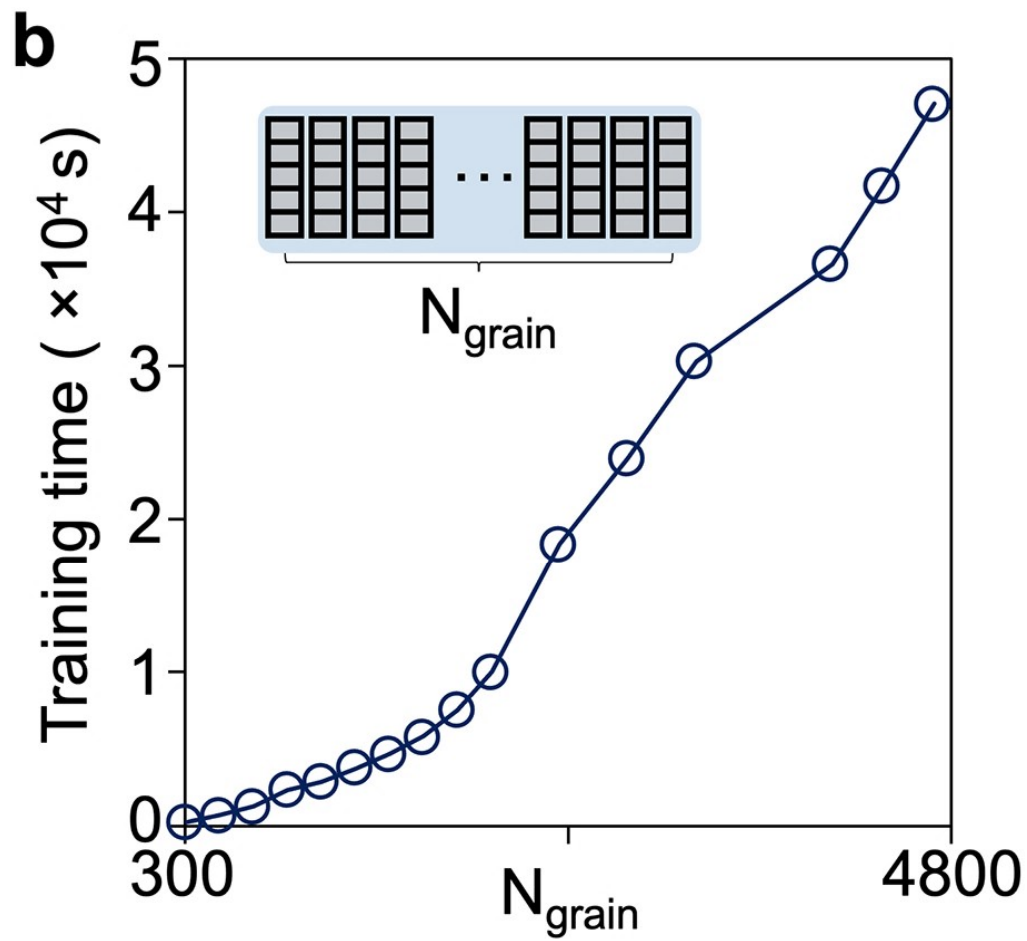

**b**

Training time ( ×10⁴ s)

$N_{grain}$

$N_{grain}$

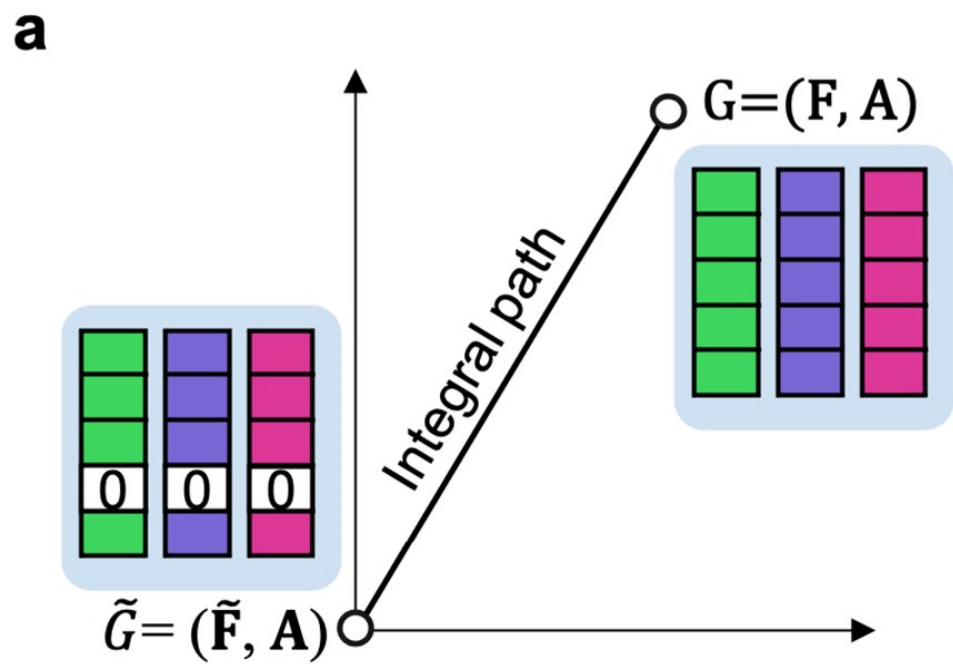

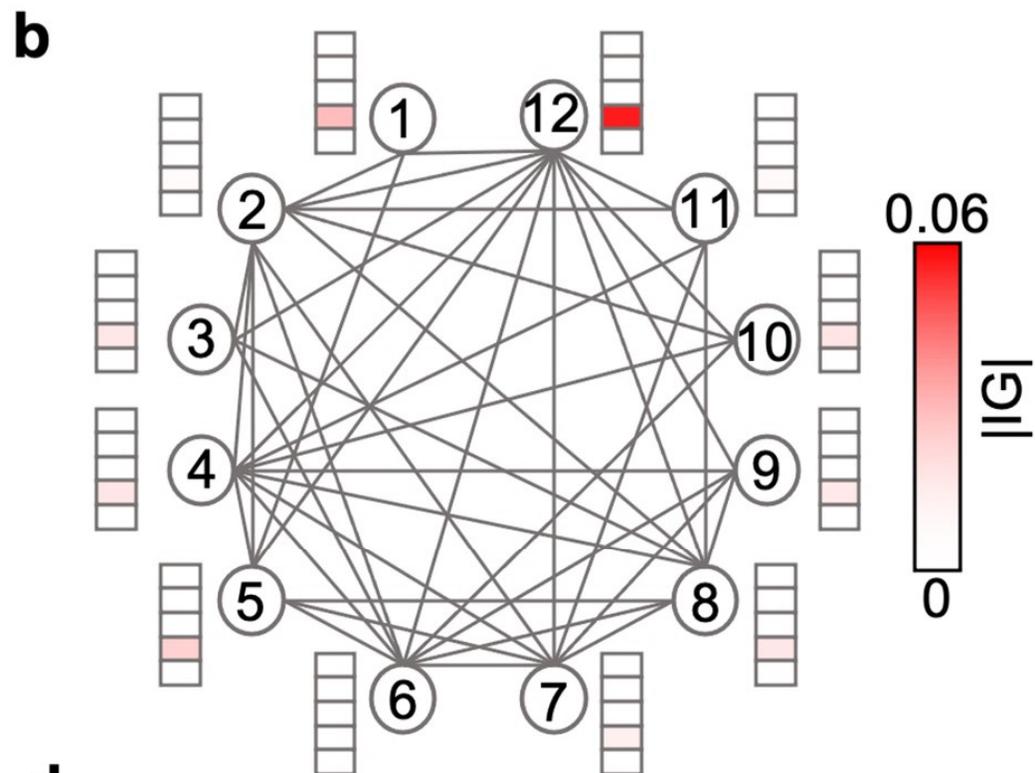

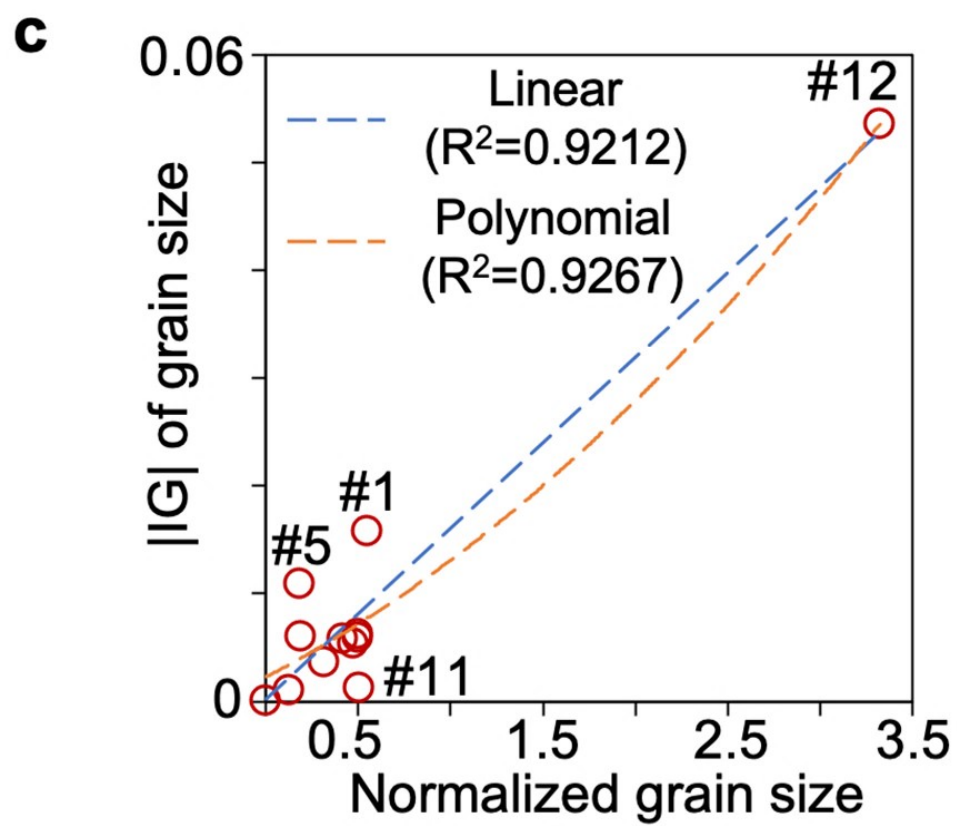

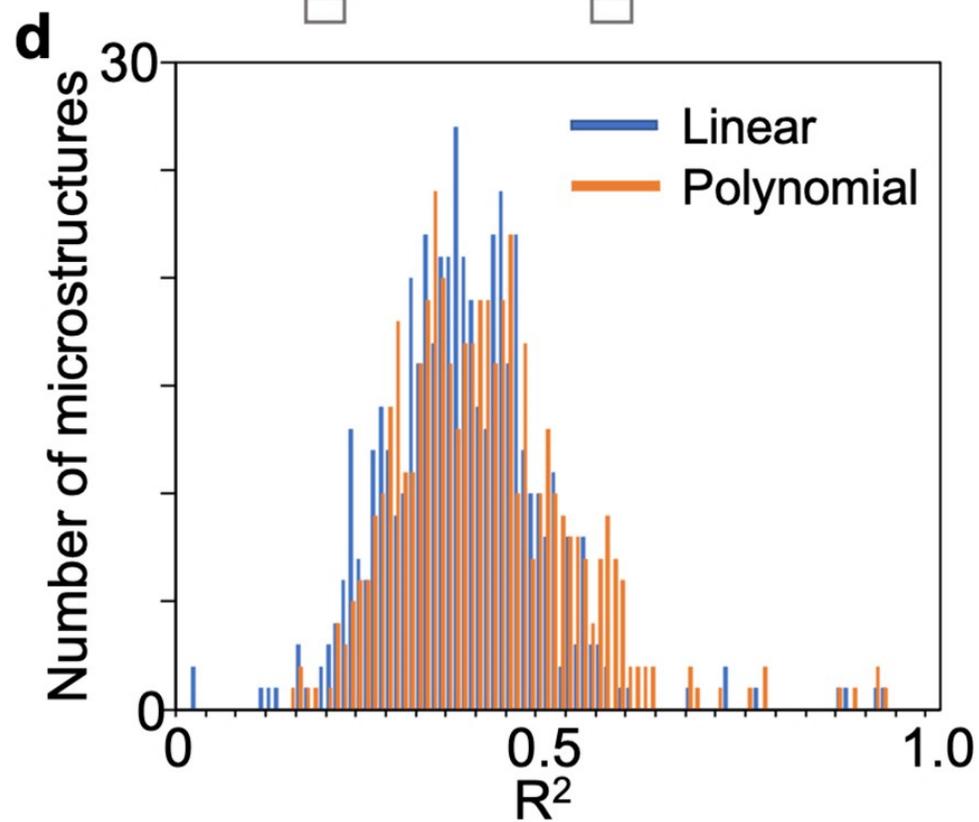